\documentstyle[12pt]{article}
\def\NPB{{\em Nucl. Phys.} B}

\def\PRD{{\em Phys. Rev.} D}

\begin{document}
\vspace*{-2cm}
\begin{flushright}
IFT 2001/10\\
\end{flushright}
\vspace*{1.cm}
\centerline{\bf\huge The prompt photon photoproduction at 
THERA~\footnote{To appear in "The THERA Book", DESY-LC-REV-2001-062}}
\vskip 2.5cm
\centerline{\Large Maria Krawczyk and Andrzej Zembrzuski}
\centerline{Institute of Theoretical Physics, Warsaw University, Poland}

\vskip 3cm
\begin{abstract}
We present NLO QCD predictions for the prompt photon photoproduction
at the THERA, and compare them with results for the HERA collider.
\end{abstract}

\section{Introduction}
Photoproduction of photons with large transverse momentum, 
$p_T\gg\Lambda_{QCD}$, in $ep$ collisions, $ep\rightarrow e\gamma X$,
is an important test of pQCD. In particular it allows to probe the 
photonic content of the photon~\cite{Bjorken:1969ja}. 
This process, called also the
prompt photon photoproduction or the Deep Inelastic Compton (DIC) scattering,
was measured at the HERA collider by the 
ZEUS~\cite{Breitweg:1997pa,unknown:1998uj,Breitweg:2000su} and H1~\cite{h1}
Collaborations.  In this note we compare the potential of the THERA and HERA
colliders in measuring the DIC process. The predictions of NLO QCD 
calculations 
for the photoproduction of {\sl non-isolated} prompt photon at HERA and 
THERA energies are presented.

In the present experimental analyses of the prompt photon production 
at HERA the 
hadronic energy detected close to the final photon has been restricted, 
and therefore one can say that the final photon is {\sl isolated}. 
The NLO calculations for the 
DIC process with isolated photon at the HERA 
collider~\cite{Gordon:1995km,Krawczyk:1998it} give 
reasonably good description of the ZEUS data for the $p_T$ 
distribution~\cite{Breitweg:2000su}. 
For the rapidity
distribution the results are in good agreement with data at 
rapidities $\eta_{\gamma}>0.1$, however they lie below data at rapidities 
$\eta_{\gamma}<0.1$ (the difference is $\sim$ 30\%)~\cite{Breitweg:2000su}. 
For 
comparison with results for non-isolated photon discussed in this
paper, we present also our 
predictions for isolated photon at HERA together with the ZEUS 
data~\cite{Breitweg:2000su}.

\section{The NLO cross section}

The cross section for the photoproduction in the $ep$ collision can
be calculated using the equivalent photon 
approximation~\cite{vonWeizsacker:1934sx}:
\begin{eqnarray}
d\sigma^{ep}=\int G_{\gamma/e}(y) 
d\sigma^{\gamma p} dy ,
\end{eqnarray}
where $y=E_{\gamma}/E_e$ is (in the laboratory frame) a fraction of the 
initial electron energy taken by the photon. 
The (real) photon distribution in the electron we take in the form:
\begin{eqnarray}
G_{\gamma/e}(y)={\alpha\over 2\pi}\{ {1+(1-y)^2\over y}
\ln [{Q^2_{max}(1-y)\over m_e^2 y^2}]\!-\!{2\over y}(1-y
-{m_e^2y^2\over Q^2_{max}})\},
\end{eqnarray}
with $m_e$ being the electron mass, and $Q^2_{max}$ = 1 GeV$^2$
(used both for the HERA and THERA collider)

We consider now the production of a large-$p_T$ photon in the 
$\gamma p\rightarrow\gamma X$ scattering.
The lowest order (Born) contribution to the cross section
comes from the Compton process on the quark,
$\gamma q\rightarrow\gamma q$. 
In the NLO calculation one takes into account 
$\alpha_s$ corrections to the Born process: the virtual gluon 
exchange and real gluon emission, together with the subprocess
$\gamma g\rightarrow\gamma q\bar{q}$. The collinear
singularities, which appear in these corrections, are subtracted and
shifted into corresponding parton densities or fragmentation function.
The remaining corrections, without singularities, constitute the 
so called K-factor.

The initial photon may interact directly with the parton from the proton 
(as in the Born process) or may interact as the resolved one via its
partons.
Analogously, the observed final photon arise directly from hard 
partonic subprocesses (as in the Born process)
or arise from fragmentation processes in which 
$q$ or $g$ 'decays' into $\gamma$.

The NLO cross section for the photon production in $\gamma p$ collision
can therefore be written in the following form~\footnote{Note, that
in our approach the parton densities in the photon and parton 
fragmentation into photon are treated as quantities of order
$\alpha_{em}$~\cite{Krawczyk:1998it}, while e.g. 
in~\cite{Gordon:1995km} they are assumed to be
of order $\alpha_{em}/\alpha_{s}$. This leads to different
set of subprocesses included in NLO calculations in both approaches.}:
\begin{eqnarray}
E_{\gamma}{d^3\sigma^{\gamma p\rightarrow\gamma X}\over d^3p_{\gamma}} = 
\sum_{q}\int dx f_{q/p}(x,\bar{Q}^2)
E_{\gamma}{d^3\sigma^{\gamma q\rightarrow \gamma q}\over d^3p_{\gamma}}+
\sum_{b}\int dx f_{b/p}(x,\bar{Q}^2) {\alpha_s(\bar Q^2)\over 
2\pi^2 \hat{s}}K_b+
\label{born}\\
+\sum_{a b}\int dx_{\gamma}\int dx f_{a/\gamma}(x_{\gamma},\bar{Q}^2) 
f_{b/p}(x,\bar{Q}^2)
E_{\gamma}{d^3\sigma^{ab\rightarrow\gamma d}\over d^3p_{\gamma}}+
\label{init}\\
+\sum_{b c}\int {dz\over z^2}\int dx f_{b/p}(x,\bar{Q}^2)
D_{\gamma /c}(z,\bar{Q}^2)
E_{\gamma}{d^3\sigma^{\gamma b\rightarrow cd}\over d^3p_{\gamma}}+
\label{final}\\
+\sum_{a b c}\int {dz\over z^2}\int dx_{\gamma}\int dx
f_{a/\gamma}(x_{\gamma},\bar{Q}^2) f_{b/p}(x,\bar{Q}^2)
D_{\gamma /c}(z,\bar{Q}^2)
E_{\gamma}{d^3\sigma^{ab\rightarrow cd}\over d^3p_{\gamma}}+
\label{double}\\
+\int dx f_{g/p}(x,\bar{Q}^2)
E_{\gamma}{d^3\sigma^{\gamma g\rightarrow \gamma g}\over d^3p_{\gamma}},
\label{box}
\end{eqnarray}
where $x_{\gamma}$ ($x$) stands for the photon (proton) momentum fraction 
taken by the $a$ ($b$)-parton,
and $z$ is the momentum fraction of the  $c$-parton fragmenting into photon.
The $f_{a/\gamma}$, $f_{b/p}$ and $D_{\gamma /c}$ are the parton densities
in the photon, parton densities in the proton and parton fragmentation
into photon, respectively.

The first and the second term in eq.~(\ref{born}) correspond to the Born 
contribution and to the K-factor, respectively. 
The three terms (\ref{init}, \ref{final}, 
\ref{double}) are due to various resolved photon subprocesses, with resolved 
initial or/and final photon.
The last term (\ref{box}) stands for the contribution of
the box diagram, $\gamma g\rightarrow\gamma g$. The box diagram is 
of NNLO-type (as it is the double resolved photon processes), 
nevertheless we take it into account in our NLO calculation,
because it is known that they give a sizable contributions, see 
e.g.~\cite{Krawczyk:1998it}.

\section{The results}
We present NLO QCD
results for the DIC cross section at HERA 
and THERA energies. 
For THERA we assume $E_e$ = 250 GeV 
and $E_p$ = 920 GeV. For HERA we take $E_e$ = 27.5 GeV and $E_p$ = 820 
GeV~\footnote{not 920 GeV} used in the
ZEUS~\cite{Breitweg:1997pa,unknown:1998uj,Breitweg:2000su} and H1~\cite{h1} 
measurements.
The GRV NLO parametrizations
for the parton distributions in the proton~\cite{Gluck:1995uf} and 
photon~\cite{Gluck:1992jc},
and parton fragmentation into photon~\cite{Gluck:1993zx} are used. The 
renormalization/factorization scale is assumed equal to the transverse 
momentum of the final photon, $\bar{Q} = p_T$. The calculations are
performed for four massless quarks, $N_f=4$, and $\Lambda_{QCD}$ = 320 MeV.

\vspace*{8.cm}
\begin{figure}[h]
\vskip 0.cm\relax\noindent\hskip -2cm
       \relax{\includegraphics{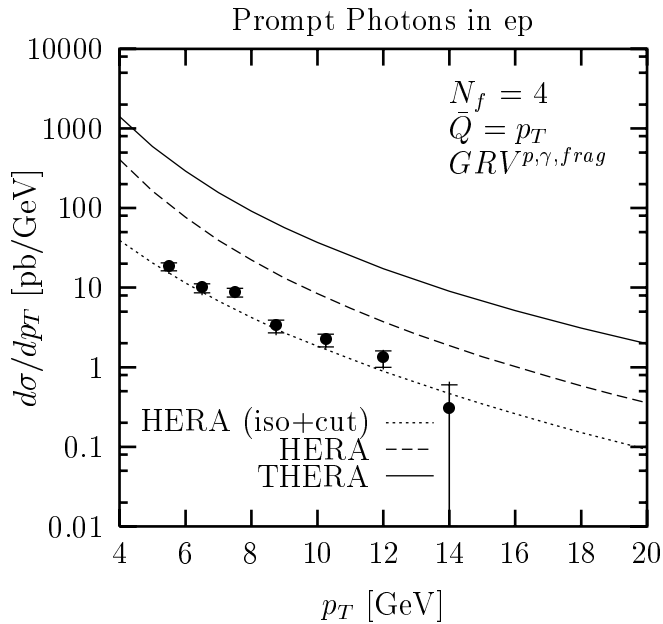}}
\vspace{0.cm}
\caption{The $p_T$ distribution for the inclusive prompt photon
production. The rapidity is taken in the range 
$-0.7\le\eta_{\gamma}\le 0.9$. The NLO results for 
THERA
(solid line) and HERA~\cite{Krawczyk:1998it} (dashed line) are shown. For
comparison also results for isolated photons 
with an additional cut ($0.2\le y\le 0.9$) as measured by the ZEUS
Collaboration~\cite{Breitweg:2000su} 
are plotted together with the NLO 
predictions~\cite{Krawczyk:1998it} (dotted line).}
\label{fig:dpt}
\end{figure}

In fig.~\ref{fig:dpt} the cross section
$d\sigma/dp_T$ is presented for 
$4\le p_T\le 20$ GeV and $-0.7\le\eta_{\gamma}\le 0.9$. 
The results strongly depend
on $p_T$: the cross section decreases by
three orders of magnitude when the $p_T$ increases from 4 to 20 GeV.
The predictions for the THERA collider are larger than for HERA, 3.5 to 5.5 
times for $p_T$ from 4 to 20 GeV. Note that the isolation and additional cuts 
applied by the ZEUS 
group~\cite{Breitweg:2000su} at HERA reduce the cross section by a 
factor of 10 (4) at $p_T=4$ (20) GeV.

\begin{figure}[h]
\vskip 0.cm\relax\noindent\hskip -2cm
       \relax{\includegraphics{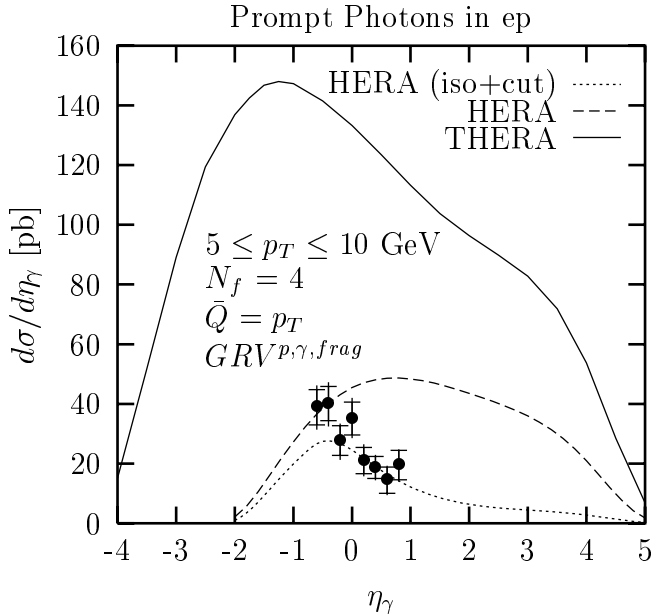}}
\vspace{0.cm}
\caption{The $\eta_{\gamma}$  distribution for $5\le p_T\le 10$ GeV.
The NLO results for THERA (solid line) and 
HERA~\cite{Krawczyk:1998it} 
(dashed line) are shown. For comparison the ZEUS Collaboration 
data~\cite{Breitweg:2000su} and the NLO 
predictions~\cite{Krawczyk:1998it} (dotted line) for isolated photons 
with additional cuts ($0.2\le y\le 0.9$, $-0.7\le\eta_{\gamma}\le 0.9$) 
are presented.}
\label{fig:dn}
\end{figure}

The cross section $d\sigma/d\eta_{\gamma}$ for $5\le p_T\le 10$ GeV
is presented in fig.~\ref{fig:dn}. The range of accessible rapidities 
is extended and the value of the cross section is much higher
for the THERA collider in comparison with predictions for HERA.
In the central rapidity region, $-1\le\eta_{\gamma}\le 1$,
(where the ZEUS data~\cite{Breitweg:2000su} are shown)
the results obtained for THERA are $\sim$ 2 - 5.5
times larger than for HERA.

\vspace*{8.cm}

\section{Summary}
We have presented results of NLO calculation for the non-isolated
prompt photon photoproduction at the THERA and HERA colliders. 
The predictions for THERA
are a few times larger than for HERA in a wide range of transverse
momentum and rapidities. This can allow to perform much more precise
measurements than the present ones and e.g. in testing the parton densities 
in the photon.
For a comparison the current data and predictions
for isolated photon at HERA are also shown. 

\vskip 0.5cm
Supported in part by Polish State Committee for Scientific
Research, grant number 2P03B05119 (2000-2001), 
and by European Commision 50th framework contract HPRN-CT-2000-00149.

\end{document}